\documentclass[11pt,graphicx]{article}
\usepackage{amssymb,amsmath,amsfonts}
\usepackage{graphicx}
\usepackage{graphics}
\usepackage{eepic,epsfig}
\usepackage{hyperref}
\usepackage{cite}

\begin{document}
\title{Vacuum current and polarization induced by magnetic flux in a higher-dimensional cosmic string in the presence of a  flat boundary}
\author{E. A. F. Bragan\c{c}a$^1$\thanks
{E-mail: eduardo.braganca@uemasul.edu.br} \, and
E. R. Bezerra de Mello$^2$\thanks{E-mail: emello@fisica.ufpb.br}
\\
\textit{$^1$Centro de Ci\^{e}ncias Agr\'{a}rias, Naturais e Letras - CCANL,}\\
\textit{Universidade Estadual da Regi\~{a}o Tocantina do Maranh\~{a}o,}\\
\textit{Avenida Brejo do Pinto S/N, 65975-000, Estreito, MA, Brazil}\vspace{0.3cm}\\
\textit{$^{2}$Departamento de F\'{\i}sica, Universidade Federal da Para\'{\i}ba,}\\
\textit{58.059-970, Caixa Postal 5.008, Jo\~{a}o Pessoa, PB, Brazil}}

	\maketitle     
	\begin{abstract}
In this present paper we investigate the vacuum bosonic current and polarization induced by a magnetic flux running along a higher dimensional cosmic string in the presence of a flat boundary orthogonal to the string. In our analysis we assume that the quantum field obeys Dirichlet or Neunmann conditions on the flat boundary. To develop this analysis we calculate the corresponding  positive frequency Wightman function. As consequence of the boundary condition, the Wightamn function is expressed in term of two contributions: The first one corresponds to the Wightman function in cosmic string spacetime in the absemce of boundary, while the second one is induced by the presence of the boundary. Due to the fact that the analysis of induced bosonic current and polarization effects in the pure cosmic string spacetime have been developed by many authors, the main objective of this paper is to study the effects induced by the boundary. Regarding to the induced current, we show that, depending on the condition adopted, the boundary-induced azimuthal current can cancel or intensifies the total induced azimuthal current on the boundary; moreover, the boundary-induced azimuthal current is a periodic odd function of the magnetic flux. As to the vacuum expectation values of the field squared and the energy-momentum tensor, the boundary-induced contributions are even functions of magnetic flux. In particular, we consider some special cases of the boundary-induced part of the energy density
and evaluate the normal vacuum force on the boundary.
		\vspace{1pc}
	\end{abstract}
\maketitle
\section{Introduction}
In the context of Grand Unified Theories, different types of
topological defects can be produced in the early Universe as consequence of
vacuum symmetry breaking phase transitions during its expansion process \cite{Kibb80,VIL,Vile94}. Depending on the topology of the vacuum manifold these defects can be  domain walls, cosmic strings, monopoles and texture. Among them cosmic strings have attracted considerable attention. Even so, on basis of recent observational data on the cosmic microwave background, the cosmic string  have been ruled out as the primary source for large scale structure formation, these topological objects are still candidates for a variety of fascinating physical phenomena \cite{Damo00,Bhat00,Bere01}. In addition,   cosmic strings have attracted a renewed interest partly because a variant of their formation mechanism is proposed in the framework of brane inflation \cite{Sara02,Cope04,Dval04}.

The geometry of the specetime produced by an idealized cosmic string, i.e., a very thin and straight linear topological defect, is characterized by a planar angle deficit in the two-dimensional sub-space orthogonal to the string. It is locally flat but presents a global conical structure. The simplest theoretical model which describes this model is given by a delta-Dirac type distribution for the energy-momentum tensor along the linear defect. Besides this idealized model, a more realistic system to describe a cosmic string comes from a classical field theory. Coupling the energy-momentum tensor associated with the vortex system investigated by Nielsen and Olesen in \cite{N-O} with the Einstein equations, Garfinkle and Linet in \cite{Garfinkle,Linet} found asymptotically cylindrical symmetric solution for the coupled system that corresponds a magnetic flux running along a very small linear tube, and a planar angle deficit in the two-surface orthogonal to it. 

 The nontrivial topology or curvature of the spacetime or even the presence of gauge field  affect the quantum vacuum associated with charged fields. Specifically the conical structure associated with an idealized cosmic string produces significant modifications on the vacuum expectation values (VEVs)  of physical observables like the energy-momentum tensor, $\langle T_{\mu\nu}\rangle$. The calculations of the VEVs of physical observables associated with the scalar and fermionic fields in the cosmic string spacetime have been developed in  \cite{PhysRevD.35.536, escidoc:153364, GL, DS, PhysRevD.46.1616,PhysRevD.35.3779, LB, BK}. Furthermore, considering  the presence of a magnetic flux running through the core of the string gives additional contributions to the VEVs associated with charged fields \cite{PhysRevD.36.3742, guim1994, SBM, SBM2, SBM3, Spinelly200477, SBM4} as well as induces vacuum current densities, $\langle j^\mu\rangle$. This phenomenon has been investigated for massless and massive scalar fields in \cite{LS} and \cite{SNDV}, respectively. In these papers, the authors have shown that induced vacuum current densities along the azimuthal direction arise if the ratio of the magnetic flux by the quantum one has a nonzero fractional part. The induced bosonic current in higher-dimensional compactified cosmic string spacetime was calculated in \cite{Braganca2015}. Moreover, the calculation of induced fermionic currents in higher-dimensional cosmic string spacetime in the presence of a magnetic flux has been developed in \cite{ERBM}. The induced fermionic current by a magnetic flux in $(2+1)$-dimensional conical spacetime and in the presence of a circular boundary has also been analyzed in \cite{PhysRevD.82.085033}. 

The presence of boundaries also produce vacuum polarization phenomena. This is the well-known Casimir effects. The analysis of Casimir effects in the idealized cosmic string space-time have been developed for scalar, fermionic and vector fields in \cite{Mello,Mello1,Aram1} obeying specific boundary conditions on cylindrical surface. Moreover, the analysis of Casimir effects induced by just one flat boundary orthogonal to the string have been developed for scalar and fermionic fields in \cite{MelloCQG2001} and \cite{MelloCQG2013}, respectively, and for the case of two flat boundaries in \cite{MelloPRD2018}\footnote{The vacuum polarization effects induced by a composite topological defect has been analyzed in \cite{Mello3}.}.
 
In this paper we want to continue in the same line of investigation of \cite{MelloCQG2001}; however, at this time considering a more general system composed by a scalar charged fields propagating in a higher dimensional cosmic string  having a magnetic flux running along its core, and considering the presence of a flat hypersurface  orthogonal to it. Two different boundaries conditions will be imposed on the field: the Dirichlet and Newman conditions. The main objective of this analysis is to investigate the influence of the boundary on the induced vacuum current $\langle j^\mu\rangle$, and on the vacuum expectation value (VEV) of the field squared, $\langle|\varphi|^2\rangle$, and the energy-momentum tensor,  $\langle T^{\mu\nu}\rangle$.

This paper is organized as follows: In Section \ref{sec2} we present the background geometry of the  spacetime that we want to work, and the explicit expression for the four-vector potential considered. Also we provide the complete set of normalized wave-function associated with a charged scalar quantum field obeying Dirichlet/Newman boundary condition on a flat plane orthogonal to the string which presents a magnetic flux along its core. By using the mode summation formula, we calculate the positive frequency Wightman function. As we will see the corresponding Wightman functions are expressed in terms of two distinct contributions: The first one is the standard Wightman function associated with a charged massive scalar field in a higher dimensional  cosmic string spacetime in absence of boundaries, and the second contribution is due to the boundary condition obeyed by the field. The first contribution is divergent at coincidence limit; as to the second one, it is finite in this limit for points away from the boundary.  In Section \ref{sec3} we investigate the vacuum bosonic current induced by the magnetic flux and boundary. There we will see that, depending on the boundary condition adopted to the field on the flat boundary, the induced azimuthal current can decreases or increases the intensity of the total induced current. In Section \ref{sec4} we calculate the contribution of the VEV of the field squared and the energy-momentum tensor induced by the boundary and magnetic flux. One of our main objective of this paper, is to investigate how the presence of a magnetic flux modifies these quantities; moreover, we will analyze these observable in different regions of the space. Specifically for points close and far from the string and/or boundary. In this sense some asymptotic expressions will be explicitly provided.  In Section \ref{conc}, we summarize the most relevant results obtained. In  this paper we will use the units $\hbar =G=c=1$.
 
\section {Wightman function}
\label{sec2}
In this section we present the geometry background of the spacetime that we want to work. It corresponds to a  generalization of a four-dimensional idealized cosmic string spacetime for higher dimensions. Considering $D\geq4$ as the dimension of the spacetime, by using cylindrical coordinate system this $D-$dimensional conical space is given by the line element below,
\begin{equation}
ds^{2}=g_{\mu\nu}dx^{\mu}dx^{\nu}=dt^{2}-dr^2-r^2d\phi^2-dz^2- \sum_{i=4}^{D-1}(dx^{i})^2 \ .
\label{eq01}
\end{equation}
In this coordinate system we are assuming: $r\geq 0$, $0\leq\phi\leq 2\pi/q$ and $-\infty< (t, \ x^i) < +\infty$ for $i=4,...,D-1$.  
The presence of the cosmic string is codified through the parameter $q> 1$. In a four dimensional spacetime, this parameter is related to the linear mass density of the string by $q^{-1}=1-4\mu $. Because we want to investigate the influence of a flat boundary orthogonal to the string located at $z=0$ on the quantum system, we will assume that the coordinate $z\geq 0$. 

The quantum dynamics of a charged bosonic field with mass $m$ in a curved spacetime and in the presence of an electromagnetic potential vector, $A_\mu$, is governed by the equation below,
\begin{equation}
\left[\frac{1}{\sqrt{|g|}}D_\mu\left(\sqrt{|g|}\,g^{\mu\nu}D_\nu\right)+m^{2}+\xi R\right] \varphi_\sigma (x)=0 \ .
\label{eq02}
\end{equation}
In the above equation we have included the non-minimal coupling between the field with the geometry, given by $\xi R$, where $R$ represents the curvature scalar, and $\xi$ the non-minimal coupling. Moreover,  $D_{\mu}=\partial_{\mu}+ieA_{\mu}$ and $g={\rm det}(g_{\mu\nu})$. 
Considering a thin and infinitely straight cosmic string, we have that $R=0$ for $r\neq 0$.

In our analysis we will assume that only the azimuthal component of the vector potential does not vanish, i.e., there is only $A_\phi={-q\Phi}/{2\pi}$, being $\Phi$ the magnetic flux along the string.

In the spacetime defined by \eqref{eq01} and in the presence of the vector 
potential given above, the equation \eqref{eq02} becomes
\begin{equation}
\left[\partial_t^2-\partial_r^2-\frac{1}{r}\partial_r-\frac{1}{r^2}(\partial_{\phi}+
ieA_{\phi})^2-\partial_{z}^2-\sum_{i=4}^{D-1}\partial_{i}^{2}+m^2\right]\varphi_\sigma(x)=0 \ . 
\label{diffe.eq}
\end{equation}
 
The positive energy solution of  this equation can be obtained  by considering the general expression,
 \begin{equation}
 \varphi_\sigma(x)=C_\sigma R(r)e^{-iE t+iqn\phi+i{\vec{k}\cdot{\vec{r}}_{\parallel}}} g(z)  \ ,
 \label{Ansatz}
 \end{equation}
 where ${\vec{r}}_{\parallel}$ represents the coordinates of the extra dimensions, $\vec{k}$ the momentum along these directions and $C_\sigma$ is a normalization constant. The unknown function $g(z)$ will be specified by the boundary condition obeyed by the field on the boundary placed at $z=0$.
 
First of all we impose that  $g(z)$ satisfies the differential equation,
 \begin{eqnarray}
 \frac{d^2g(z)}{dz^2}=-k_z^2g(z) \  .
 \label{g.function}
 \end{eqnarray} 
Accepting that, and substituting \eqref{Ansatz} into \eqref{diffe.eq}, the differential equation for the radial function $R(r)$ becomes, 
\begin{equation}
\left[\frac{d^2}{dr^2}+\frac{1}{r}\frac{d}{dr}+\lambda^2-\frac{q^2(n+\alpha)^2}{r^2}\right]R(r)=0 \ , 
\label{eq.radial}
\end{equation}
where 
\begin{eqnarray}
\lambda&=&\sqrt{E^2-{\vec{k}}^{2}-k_z^2-m^2} \ , \nonumber\\  
\alpha&=&-\frac{e\Phi}{2\pi} \ .
\label{def1}
\end{eqnarray}

The solution of \eqref{eq.radial} regular at $r=0$, is:
\begin{eqnarray}
R(r)=J_{q|n+\alpha|}(\lambda r) \  ,
\label{Rsolution}
\end{eqnarray}
being $J_\nu(z)$ the Bessel function \cite{grad}. As to the solution of \eqref{g.function}, we have two possibilities:
\begin{itemize}
\item For the field obeying Dirichlet condition, $g(z)=0$ at $z=0$, we have
\begin{equation}
g(z)=\sin(k_z z) \  ,
\label{D.cond}
\end{equation}
\item For the field obeying Newman condition, $\frac{dg(z)}{dz}=0$ at $z=0$, we havve
\begin{equation}
g(z)=\cos(k_z z) \  .
\label{N.cond}
\end{equation}
\end{itemize}

The solution \eqref{Ansatz} is then characterized by the set of quantum number, $\sigma=\{\lambda, \ n, \ k_z, \ k^i \}$. Its normalization constant $C_\sigma$ can be obtained by the normalization condition
\begin{equation}
i\int d^Dx\sqrt{|g|}\left[\varphi_{\sigma'}^{*}(x)\partial_t
\varphi_{\sigma}(x)-\varphi_{\sigma}(x)\partial_t\varphi_{\sigma'}^{*}(x)
\right]=\delta_{\sigma,\sigma'} \ ,
\label{norm}
\end{equation}
where the delta symbol on the right-hand side is understood as Dirac delta 
function for the continuous quantum number, $\lambda$, $k_z$ and ${\vec{k}}$, and Kronecker delta for the discrete ones, $n$. From \eqref{norm} one finds 
\begin{equation}
|C_\sigma|=\sqrt{\frac{2q\lambda}{(2\pi)^{D-2}E}} \ ,
\label{constC}
\end{equation}
for both modes of wave-function $g(z)$.

The properties of the vacuum state can be described in terms of the positive frequency Wightman function, $W(x,x')=\left\langle 0|\varphi(x)\varphi^{*}(x')|0 \right\rangle$, where $|0 \rangle$ represents the vacuum state. Having this function we can evaluate the induced bosonic current and the VEV of the field squared and energy-momentum tensor. 

For the evaluation of the Wightman function, we adopt the mode sum formula
\begin{equation}
W(x,x')=\sum_{\sigma}\varphi_{\sigma}(x)\varphi_{\sigma}^{*}(x') \ ,
\label{W.function}
\end{equation}
where we are using the compact notation for the sum  defined as
\begin{equation}
\sum_{\sigma }=\sum_{n=-\infty}^{+\infty} \ \int d{\vec{k}} \ \int_0^\infty \ d k_z \int_0^\infty
\ d\lambda  \ .  \label{Sumsig}
\end{equation}
The set $\{\varphi_{\sigma}(x), \ \varphi_{\sigma}^{*}(x')\}$ represents a complete set of normalized mode functions satisfying the Dirichlet/Newman condition at the boundary.

Substituting \eqref{Ansatz} with \eqref{Rsolution} and \eqref{constC} into \eqref{W.function}, we obtain:
\begin{eqnarray}
W(x,x')&=&\frac{2q}{(2\pi)^{D-2}}\sum_{n=-\infty}^\infty\int_0^\infty d\lambda \lambda J_{q|n+\alpha|}(\lambda r)J_{q|n+\alpha|}(\lambda r')e^{iqn(\phi-\phi')}\nonumber\\
&\times&\int_0^\infty dk_z g(z)g(z')\int d^{D-4}\vec{k} e^{i{\vec{k}}\cdot({\vec{r}}_{\parallel}-{\vec{r'}}_{\parallel})}\frac{e^{-iE(t-t')}}{\sqrt{\lambda^2+k_z^2+{\vec{k}}^2+m^2}}  \  .
\label{W.function1}
\end{eqnarray}

By adopting a Wick rotation, $t\to -i\tau$ and using the identity below,
\begin{eqnarray}
\frac{e^{-E\Delta\tau}}{E}=\frac2{\sqrt{\pi}} \int_0^\infty ds e^{-E^2s^2-\Delta\tau^2/(4s^2)}  \  , 
\label{ident1}
\end{eqnarray}
we obtain,
\begin{eqnarray}
W(x,x')&=&\frac{4q}{(2\pi)^{D-2}\sqrt{\pi}}\sum_{n=-\infty}^\infty\int_0^\infty ds \int_0^\infty d\lambda \lambda e^{-\lambda^2s^2}J_{q|n+\alpha|}(\lambda r)J_{q|n+\alpha|}(\lambda r')e^{iqn(\phi-\phi')}\nonumber\\
&\times&\int_0^\infty dk_z e^{-k_z^2s^2} g(z)g(z')\int d^{D-4}\vec{k} e^{-\vec{k}^2s^2} e^{i{\vec{k}}\cdot({\vec{r}}_{\parallel} -{\vec{r'}}_{\parallel})}e^{-m^2s^2}e^{-\Delta\tau^2/(4s^2)} \ .
\label{W.function2}
\end{eqnarray} 
For the integration over the quantum number $\lambda$ we use the formula \cite{grad}
\begin{equation}
\int_0^\infty d\lambda \ \lambda J_{\beta}(\lambda r)J_{\beta}(\lambda r')e^{-s^2\lambda^2}=\frac{1}{2s^2}
\mathrm{exp}\left(-\frac{r^2+r'^2}{4s^2}\right)I_{\beta}\left(\frac{rr'}{2s^2}\right) \  .
\label{int_lambda}
\end{equation}
As to the integral in $k_z$ we have,
\begin{eqnarray}
\int_0^\infty e^{-k_z^2 s^2} g(z)g(z') dk_z= \frac{\sqrt{\pi}}{4s}\left(e^{-\frac{(z-z')^2}{4s^2}}\mp e^{-\frac{(z+z')^2}{4s^2}}\right) \  .
\end{eqnarray}
The negative/positive signal in the expression above refers to Dirichlet/Newman boundary condition.

So, the positive energy Wightman function given by \eqref{W.function2} can be express in terms of two function, as exhibit below:
\begin{eqnarray}
\label{Wfunc0}
W(x,x')=W_{cs}(x,x')\mp W_b(x,x') \ .
\end{eqnarray}
The negative/positive signal corresponds to the Dirichlet/Newman condition. 

Defining a new variable $u=1/2s^2$, from \eqref{W.function2} we find
\begin{eqnarray}
W_{cs}(x,x')&=&\frac{q}{2(2\pi)^{D/2}}\int_0^\infty du \ u^{\frac{D}{2}-2}e^{-\frac{m^2}{2u}-\frac{u}{2}{\cal{V}}^2_{(-)}} \ \mathcal{I}(\alpha,\Delta\phi,urr')  \  ,
\label{Wcs}
\end{eqnarray}
and
\begin{eqnarray}
W_{b}(x,x')&=&\frac{q}{2(2\pi)^{D/2}}\int_0^\infty du \ u^{\frac{D}{2}-2}e^{-\frac{m^2}{2u}-\frac{u}{2}{\cal{V}}^2_{(+)}} \ \mathcal{I}(\alpha,\Delta\phi,urr')  \  ,
\label{Wb}
\end{eqnarray}
with
\begin{eqnarray}
{\cal{V}}^2_{(\mp)}=r^2+r'^2+|\Delta\vec{r}_{||}|^2+(z\mp z')^2- (\Delta t)^2 \ .
\end{eqnarray}
Moreover, we have introduced the notation
\begin{equation}
\mathcal{I}(\alpha,\Delta\phi,urr')=\sum_{n=-\infty}^\infty e^{inq\Delta\phi}I_{q|n+\alpha|}(urr') \  .
\label{nsum}
\end{equation}

We can obtain a more convenient expressions for \eqref{Wcs} and \eqref{Wb} writing the parameter $\alpha$  defined in \eqref{def1} in the form
\begin{equation}
\alpha=n_0 +\alpha_0 \ \ {\rm with  \ |\alpha_0| \ < 1/2},
\label{alphazero}
\end{equation}
where $n_0$ is an integer number. Using the form of the summation over $n$ given in \cite{deMello:2014ksa}, we get
\begin{eqnarray}
\mathcal{I}(\alpha,\Delta\phi,x)&=&\frac{1}{q}\sum_k e^{x\cos(2k\pi /q-\Delta\phi)}e^{i\alpha(2k\pi -q\Delta\phi)}-
\frac{e^{-iqn_0\Delta\phi}}{2\pi i}\sum_{j=\pm 1}je^{ji\pi q\alpha_0}\nonumber\\
&\times&\int_0^\infty dy \frac{\cosh[qy(1-\alpha_0)]-\cosh(q\alpha_0y)e^{-iq(\Delta\phi+j\pi)})}{e^{x\cosh y}[\cosh(qy)-\cos(q(\Delta\phi+j\pi))]}.
\label{representation}
\end{eqnarray}
As to the summation over $k$ there exist the condition
\begin{equation}
-\frac{q}{2}+\frac{2\pi}{q}\Delta\phi \leq k \leq \frac{q}{2}+\frac{2\pi}{q}\Delta\phi.
\label{conditionk}
\end{equation}

Substituting \eqref{representation} with $x=urr'$ into the \eqref{Wcs} and \eqref{Wb} we get:
\begin{eqnarray}
W_{cs}(x,x')&=&\frac{m^{D-2}}{(2\pi)^{\frac{D}{2}}}\left\{\sum_k e^{i\alpha(2k\pi-q\Delta\phi)}
f_{\frac{D-2}{2}}\left[m\sqrt{{\cal{V}}_{(-)}^2-2rr'\cos(2k\pi/q-\Delta\phi)}\right]\right.\nonumber\\
&-&\left.\frac{qe^{-iqn_0\Delta\phi}}{2\pi i}\sum_{j=\pm1}je^{ij\pi q\alpha_0}\int_0^\infty dy
\frac{\cosh[qy(1-\alpha_0)]-\cosh(q\alpha_0y)e^{-iq(\Delta\phi+j\pi)}}{\cosh(qy)-\cos[q(\Delta\phi+j\pi)]}\right.\nonumber\\
&\times&\left. f_{\frac{D-2}{2}}\left[m\sqrt{{\cal{V}}_{(-)}^2+2rr'\cosh y}\right]
\right\}
\label{wcs1}
\end{eqnarray}
and
\begin{eqnarray}
W_{b}(x,x')&=&\frac{m^{D-2}}{(2\pi)^{\frac{D}{2}}}\left\{\sum_k e^{i\alpha(2k\pi-q\Delta\phi)}
f_{\frac{D-2}{2}}\left[m\sqrt{{\cal{V}}_{(+)}^2-2rr'\cos(2k\pi/q-\Delta\phi)}\right]\right.\nonumber\\
&-&\left.\frac{qe^{-iqn_0\Delta\phi}}{2\pi i}\sum_{j=\pm1}je^{ij\pi q\alpha_0}\int_0^\infty dy
\frac{\cosh[qy(1-\alpha_0)]-\cosh(q\alpha_0y)e^{-iq(\Delta\phi+j\pi)}}{\cosh(qy)-\cos[q(\Delta\phi+j\pi)]}\right.\nonumber\\
&\times&\left. f_{\frac{D-2}{2}}\left[m\sqrt{{\cal{V}}_{(+)}^2+2rr'\cosh y}\right]
\right\},
\label{wb1}
\end{eqnarray}
In the above equations we have used the definition
\begin{equation}
f_\nu (x)=\frac{K_\nu (x)}{x^\nu}.
\label{def.f}
\end{equation}

At this point the results obtained deserve to be commented: The Wightman function \eqref{wcs1} is divergent at the coincidence limit and its divergence comes from the term $k=0$. As to \eqref{wb1}, it is a consequence of the boundary condition imposed on the field. This function is
finite at coincidence limit for points outsides the boundary.

\section{Current densities}
\label{sec3}
The bosonic current density operator is given by
\begin{eqnarray}
j_{\mu }(x)&=&ie\left[\varphi ^{*}(x)D_{\mu }\varphi (x)-
(D_{\mu }\varphi)^{*}\varphi(x)\right] \nonumber\\
&=&ie\left[\varphi^{*}(x)\partial_{\mu }\varphi (x)-\varphi(x)
(\partial_{\mu }\varphi(x))^{*}\right]-2e^2A_\mu(x)|\varphi(x)|^2 \   .
\label{J.mu}
\end{eqnarray}
Its vacuum expectation value (VEV) can be evaluated in terms of the positive frequency Wightman function as shown below:
\begin{equation}
\left\langle j_{\mu}(x) \right\rangle=ie\lim_{x'\rightarrow x}
\left\{(\partial_{\mu}-\partial_{\mu '})W(x,x')+2ieA_\mu W(x,x')\right\} \ .
\label{current}
\end{equation}
However, for the case under consideration, the only nonzero component of the current density is the azimuthal one. In this way,
we will focus only on the evaluation of this component.

So, specifically the azimuthal component, reads:
\begin{equation}
\left\langle j_{\phi}(x) \right\rangle = ie \lim_{x '\rightarrow x}
\left\{(\partial_{\phi}-\partial_{\phi '})W(x,x')+2iq\alpha W(x,x')\right\} \ ,
\label{jphi}
\end{equation}
where we have substitute $A_\phi=q\alpha/e$.

Because the Wightman function \eqref{Wfunc0} is expressed as the sum of two contributions, the azimuthal current can be expressed as:
\begin{eqnarray}
\label{jphi_total}
\left\langle j_{\phi}(x) \right\rangle=\left\langle j_{\phi}(x) \right\rangle_{cs}\mp\left\langle j_{\phi}(x) \right\rangle_b \ .
\end{eqnarray}
The first contribution corresponds to the induced current in the cosmic string spacetime in absence of boundary while the second one is induced by the boundary. The latter can be negative or positive, depending on the boundary condition obeyed by the scalar field. Because the $\left\langle j_{\phi}(x) \right\rangle_{cs}$ has been calculated by many authors, here we are mainly interested in the analysis of $\left\langle j_{\phi}(x) \right\rangle_b$.

Returning to \eqref{Wb} and defining a dimensionless variable $w=ur^2$, from \eqref{jphi}, we get:
\begin{eqnarray}
\label{jphi1}
\left\langle j_{\phi}(x) \right\rangle_b=-\frac{eq}{(2\pi)^{D/2}}\frac1{r^{D-2}}\int_0^\infty dw w^{D/2-2}e^{-w(1+2(z/r)^2)}e^{-m^2r^2/(2w)}I(q,\alpha, w) \ ,
\end{eqnarray}
where 
\begin{eqnarray}
\label{Isum}
I(q,\alpha,w)=\sum_{n=-\infty}^\infty q(n+\alpha)I_{q|n+\alpha|}(w) \  .
\end{eqnarray}
In \cite{Braganca2015} we have obtained a compact expression for the summation above:
\begin{eqnarray}
I(q,\alpha,w)=\frac{2w}{q}\sideset{}{'}\sum_{k=1}^{[q/2]}\sin(2k\pi/q)\sin(2k\pi\alpha_0)
e^{w\cos(2k\pi/q)}+\frac{w}{\pi}\int_{0}^{\infty}dy\sinh y
\frac{e^{-w\cosh y}g(q,\alpha_0,y)}{\cosh(qy)-\cos(\pi q)} \  ,
\label{Isum1}
\end{eqnarray}
with
\begin{eqnarray}
g(q,\alpha_0,y)=\sin(q\pi\alpha_0)
\sinh[(1-|\alpha_0|)qy]-\sinh(yq\alpha_0)\sin[(1-|\alpha_0|)\pi q]  \  .
\label{gfunc}
\end{eqnarray}
In \eqref{Isum1}, the symbol $[q/2]$ represents the integer part of $q/2$, and the prime on the sign of the summation means that in the case $q=2p$ the term $k=q/2$ should be taken with the coefficient $1/2$.

Substituting \eqref{Isum1} and \eqref{gfunc} into \eqref{jphi1}, and using  the integral representation below for the Macdonald function \cite{grad}
\begin{equation}
K_{\nu}(x)=\frac{1}{2}\left(\frac{x}{2}\right)^{\nu}\int_{0}^{\infty}
dt\frac{e^{-t-\frac{x^{2}}{4t}}}{t^{\nu +1}} \ , 
\label{Macdonald_repres}
\end{equation}
after some intermediate steps we obtain:
\begin{eqnarray}
\langle j^{\phi}(x) \rangle_{b}&=&\frac{4em^{D}}{(2\pi)^{\frac{D}{2}}}
\left[\sideset{}{'}\sum_{k=1}^{[q/2]}\sin(2k\pi/q)\sin(2k\pi\alpha_{0})\,
f_{\frac{D}{2}}\left(\gamma_{k,z}\right)\right.\nonumber\\
&+&\left.\frac{q}{\pi}
\int_{0}^{\infty}dy \ \frac{g(q,\alpha_{0},2y)\sinh (2y)}{\cosh(2qy)-\cos(q\pi)}\,f_{\frac{D}{2}}\left(\gamma_{k,y}\right)
\right]  \  ,
\label{jphi_b}
\end{eqnarray}
where
\begin{equation}
\gamma_{k,z}=2mr\sqrt{s_k^2+(z/r)^2} \quad {\rm and} \quad
\gamma_{y,z}=2mr\sqrt{c_y^2+(z/r)^2},
\end{equation}
with
\begin{eqnarray}
s_k=\sin(k\pi/q) \   \ {\rm and} \  c_y=\cosh(y) \  .
\end{eqnarray}

The azimuthal current induced by the boundary is an odd function of $\alpha_0$. In addition, we can note that $\langle j^{\phi}(x) \rangle_{b}$  is finite in $r=0$ for points outside the boundary if $q|\alpha_0|>1$. In the case where $q|\alpha_0|<1$ the azimuthal current diverges and the leading divergence term is
given by
\begin{equation}
\langle j^{\phi}(x) \rangle_{b}\approx \frac{4em^Dq\sin(q\pi\alpha_0)}{\pi^{\frac{D+1}{2}}(2mr)^{2(1-q|\alpha_0|)}}\,
f_{\frac{D-3}{2}+2q|\alpha_0|}\left(2^{3/2}mz\right),
\label{div.J}
\end{equation}
where we have used the notation \eqref{def.f}.
Moreover, for the case of the field obeying the Dirichlet boundary condition the total current, \eqref{jphi_total}, vanishes on the boundary. Also we can see that the
azimuthal current density goes exponentially to zero for large distance from the boundary, i.e. $z/r\gg1$, according with
\begin{eqnarray}
\langle j^{\phi}(x) \rangle_{b} &\approx& \frac{e}{z^{\frac{D+1}{2}}}\left(\frac{m}{4\pi}\right)^{\frac{D-1}{2}}
\left[\sideset{}{'}\sum_{k=1}^{[q/2]}\sin(2k\pi/q)\sin(2k\pi\alpha_{0})
e^{-2mz}\right.\nonumber\\
&+&\left.\frac{q}{\pi} \int_{0}^{\infty}dy \ \frac{g(q,\alpha_{0},2y)\sinh (2y)}{\cosh(2qy)-\cos(q\pi)}
\frac{e^{-\gamma_{k,y}}}{[1+(rc_y/z)^2]^{(D+1)/4}} \right]  \  .
\label{jphi_b1}
\end{eqnarray}
The massless limit of $\langle j^{\phi}(x) \rangle_{b}$ can be obtained by taking the asymptotic limit of modified Bessel function for small arguments \cite{abramo}. In this limit, from \eqref{def.f} we have,
\begin{eqnarray}
f_\nu(z)= 2^{\nu-1}\frac{\Gamma(\nu)}{z^{2\nu}} \  .
\label{bessel-massless}
\end{eqnarray}
So using the above result we obtain:
\begin{eqnarray}
\langle j^{\phi}(x) \rangle_{b} &=& \frac{e\Gamma(D/2)}{2^{D-1}\pi^{D/2}r^D}
\left[\sideset{}{'}\sum_{k=1}^{[q/2]}\frac{\sin(2k\pi/q)\sin(2k\pi\alpha_{0})}{[s_k^2+(z/r)^2]^{D/2}}
\right.\nonumber\\
&+&\left.\frac{q}{\pi} \int_{0}^{\infty}dy \ \frac{g(q,\alpha_{0},2y)\sinh (2y)}{\cosh(2qy)-\cos(q\pi)} \frac 1{[c_y^2+(z/r)^2]^{D/2}}\right]  \  .
\label{jphi_b2}
\end{eqnarray}

\begin{figure}[h]
	\centering
	{\includegraphics[width=0.49\textwidth]{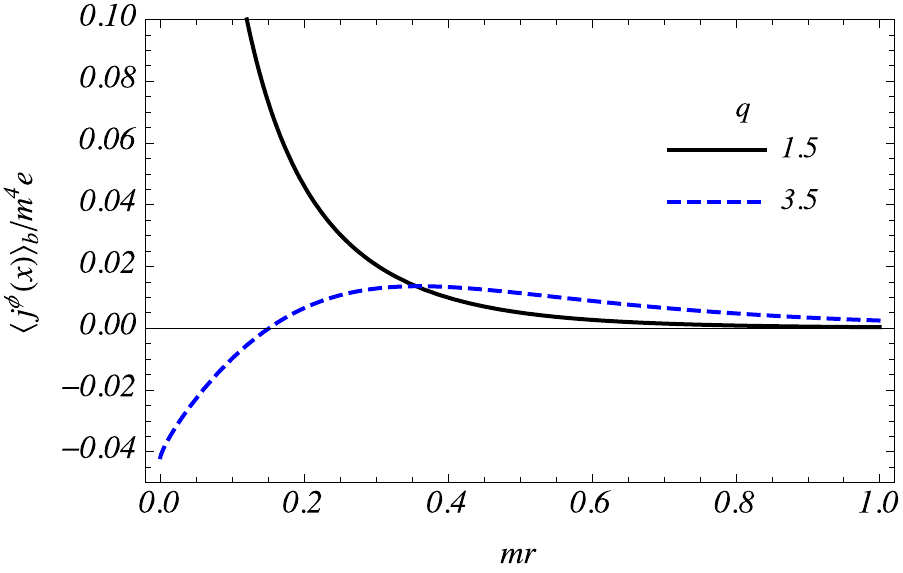}}
	\hfill
	{\includegraphics[width=0.48\textwidth]{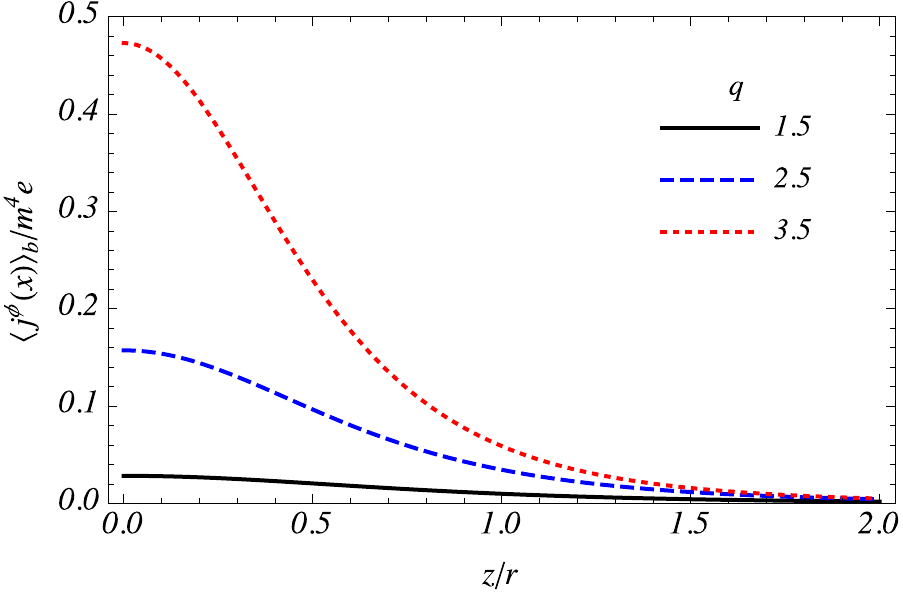}}
	\caption{VEV of the azimuthal current density induced by the boundary as function of $mr$ (left plot) and $z/r$ (right plot)
	for different values of the parameter $q$ and $D=4$.  In addition,  in the left plot we consider $mz=0.5$ and $\alpha_0=0.4$, while in
	the right one we have $mr=1$ and $\alpha_0=0.25$.}
	\label{fig01}
\end{figure}

In the Fig. \ref{fig01} we show the behavior of the VEV of the azimuthal current density as function of $mr$ (left plot) and
$z/r$ (right plot). We note that as $mr$ increase, $\langle j^{\phi}(x) \rangle_{b}$ goes to zero. On the other hand the behavior of the azimuthal current near the string,  depends crucially on the values of the product $q|\alpha_0|$. In this sense the current, as we have shown in  \eqref{div.J}, can be finite or divergent. The left plot exhibit explicitly this characteristic. Also, as $z/r$ increases $\langle j^{\phi}(x) \rangle_{b}$ goes to zero, which agrees with the Eq. \eqref{jphi_b1}. Another important point that is not evident in \eqref{jphi_b2}, and deserves to be analyzed, is the dependence of the current with $q$. In order to see that $\langle j^{\phi}(x) \rangle_{b}$ is numerically evaluated for different value of $q$ in the right plot. As we can see the intensity of the current increases when we increase $q$.

\section{Vacuum polarization}
\label{sec4}
In this section we want to develop the calculations of two important characteristics of the vacuum state: the VEVs of the
the field squared, $\langle|\varphi(x)|^2\rangle$, and the energy-momentum tensor, $\langle T_{\mu\nu}(x)\rangle$. Let us begin
with the VEV of the field squared.

\subsection{Calculation of $\langle|\varphi(x)|^2\rangle$}
\label{sec4a}
By taking into account the Eq. \eqref{Wfunc0}, the VEV of the field squared can be decomposed in the same way. Here, we are mainly
interested in the effects induced by the boundary. So, we will consider only the analysis of the VEV of the field squared induced by it.
This part of the field squared, can be obtained by taking the limit of coincidence in \eqref{wb1} as follows
\begin{equation}
\langle|\varphi(x)|^2\rangle_b=\lim_{x'\rightarrow x}W_b(x,x'),
\end{equation}
After taking the coincidence limit and solve the summation over $j$, we can write the boundary-induced part of the field squared as
\begin{equation}
\langle|\varphi(x)|^2\rangle_b=\langle|\varphi(x)|^2\rangle_b^{(M)}+\langle|\varphi(x)|^2\rangle_b^{(q,\alpha_0)},
\label{decomp_phi}
\end{equation}
where
\begin{equation}
\langle|\varphi(x)|^2\rangle_b^{(M)}=\frac{m^{D-2}}{(2\pi)^{D/2}}\,f_{\frac{D-2}{2}}\left(2mz\right).
\label{fieldsquaredMink}
\end{equation}
The above expression is the $k=0$ term of \eqref{wb1} with the coefficient $1/2$ and corresponds to the one for a boundary in Minkowski spacetime in the
absence of the cosmic string and magnetic flux.
Note that this contribution also is independent of
the radial coordinate. The second term on the right-hand side of \eqref{decomp_phi} is the contribution induced by the conical geometry, boundary and the magnetic flux. It is given by
\begin{eqnarray}
\langle|\varphi(x)|^2\rangle_b^{(q,\alpha_0)}&=&\frac{2m^{D-2}}{(2\pi)^{D/2}}
\left\{\sum_{k=1}^{[q/2]}\cos(2k\pi\alpha_0)\,f_{\frac{D-2}{2}}\left(\gamma_{k,z}\right)\right.\nonumber\\
&&\left. -\frac{q}{\pi}\int_0^\infty dy\frac{h(q,\alpha_0,2y)}{\cosh(2qy)-\cos(q\pi)}\,f_{\frac{D-2}{2}}\left(\gamma_{k,y}\right)\right\},
\label{fieldsquared}
\end{eqnarray}
with
\begin{equation}
h(q,\alpha_0,2y)=\cosh[2qy(1-|\alpha_0|)]\sin(q\pi|\alpha_0|)+\cosh(2q\alpha_0y)\sin[q\pi(1-|\alpha_0|)].
\label{hfunction}
\end{equation}
In the interval where $1\leq q<2$, the first term in the square bracket of the Eq. \eqref{fieldsquared} is absent. Note that the field squared is
an even function of $\alpha_0$.

The boundary induced part of the field squared can be considered in special cases. In the regions near the boundary, $m|z|\ll 1$ and $|z|\ll r$, the leading contribution is due to the VEV \eqref{fieldsquaredMink} being given by
\begin{equation}
\langle|\varphi(x)|^2\rangle_b^{(M)}\approx \frac{\Gamma(D/2-1)}{(4\pi)^{D/2}|z|^{D-2}}.
\label{FieldSmallZ}
\end{equation}
Considering a massless scalar field, by taking into
account the Eq. \eqref{bessel-massless}, the field squared is given by
\begin{eqnarray}
\langle|\varphi(x)|^2\rangle_b^{(q,\alpha_0)}&=&\frac{\Gamma(D/2-1)}{2^{D-1}\pi^{D/2}r^{D-2}}
\left\{\sum_{k=1}^{p}\frac{\cos(2k\pi\alpha_0)}{\left[s_k^2+(z/r)^2\right]^{D/2-1}}\right.\nonumber\\
&&\left. -\frac{q}{\pi}\int_0^\infty dy\frac{h(q,\alpha_0,2y)}{\cosh(2qy)-\cos(q\pi)}\left[c_y^2+(z/r)^2\right]^{1-D/2}\right\}.
\label{fieldsquared_massless}
\end{eqnarray}
In the regime where $z/r\gg 1$, we have
\begin{eqnarray}
\langle|\varphi(x)|^2\rangle_b^{(q,\alpha_0)}&\approx&\frac{m^{\frac{D-3}{2}}}{(4\pi z)^{\frac{D-1}{2}}}
\left\{\sum_{k=1}^{p}\cos(2k\pi\alpha_0)e^{-2mz}
-\frac{q}{\pi}\int_0^\infty dy\frac{h(q,\alpha_0,2y)e^{-\gamma_{k,y}}}{\cosh(2qy)-\cos(q\pi)}\right\}.\nonumber\\
\label{fieldLargeZ}
\end{eqnarray}

\begin{figure}[h]
	\centering
	{\includegraphics[width=0.48\textwidth]{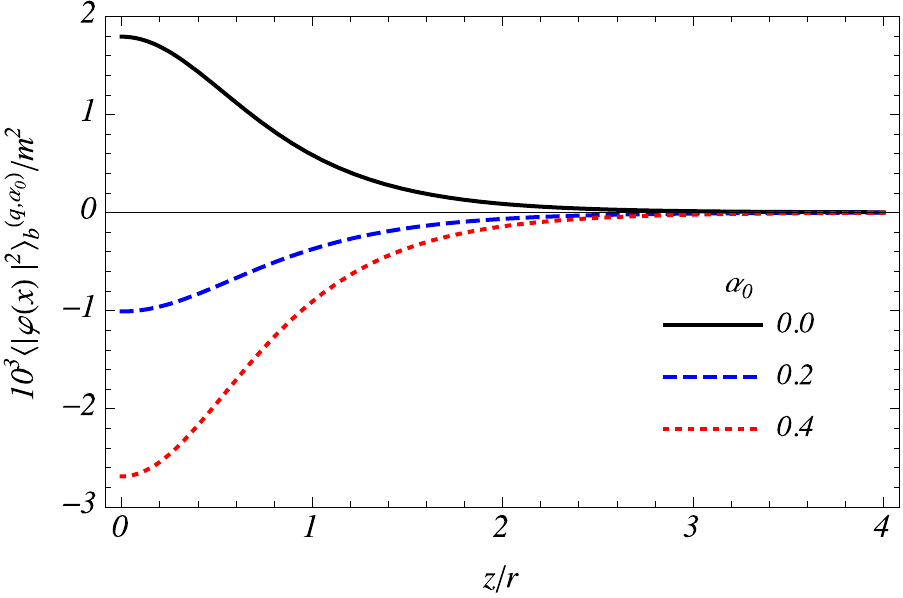}}
	\hfill
	{\includegraphics[width=0.49\textwidth]{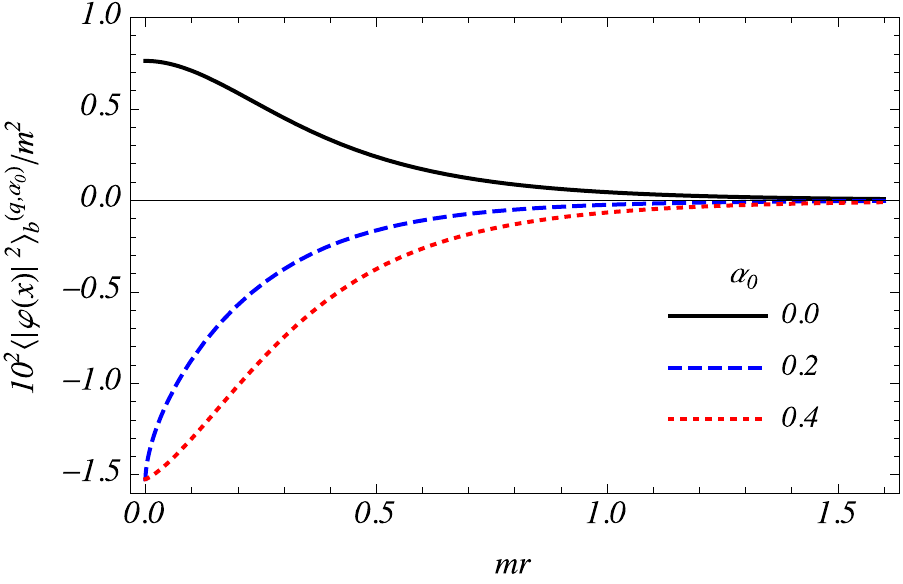}}
	\caption{VEV of the field squared as function of $z/r$ (left plot) and $mr$ (right plot) considering different values of $a_0$ and
	$D=4$. We also considered $mr=0.75$ in the left plot and $mz=0.5$ in the right one. And for both plots we have $q=1.5$.}
	\label{fig02}
\end{figure}

The behavior of the boundary induced part of the field squared is shown in Fig. \ref{fig02} as function of $z/r$ (left plot) and
as function of $mr$ (right plot) considering different values of the magnetic flux and $q=1.5$.

\subsection{Calculation of $\langle T_{\mu\nu}(x)\rangle$}
\label{sec4b}
Another quantity which characterizes the quantum state is the VEV of the energy momentum-tensor. Here, we are interested mainly in the boundary effects.
So, as for the others physical observables in this paper, we will determine only the contribution of the energy-momentum tensor
induced by the boundary. In order to evaluate this VEV, we use
the following formula obtained in \cite{Wagner}
\begin{equation}
\langle T_{\mu\nu}(x)\rangle_b=\lim_{x'\rightarrow x}\left(D_{\mu}D_{\nu '}^{*}+D_{\mu '}^{*}D_{\nu }\right)W_b(x,x')-
2\left[\xi R_{\mu\nu}+\xi \nabla_\mu \nabla_\nu-(\xi-1/4)g_{\mu\nu}\Box\right]\langle|\varphi^2(x)|\rangle_b.
\label{EMT}
\end{equation}
In the spacetime that we are considering here, the Ricci tensor for points outside the string vanishes.

We start considering the d'Alembertian operator of the field squared. This operator presents a dependence only on
the radial and axial coordinates. As the field squared is decomposed into two contributions, we also have two contributions
for the d'Alembartian operator of it.  These contributions are given by
\begin{align}
\Box\langle|\varphi(x)|^2\rangle_b^{(M)}&=-\frac{m^D}{(2\pi)^{D/2}}\left[(2mz)^2\, f_{\frac{D+2}{2}}(2m|z|)-f_{\frac{D}{2}}(2m|z|)\right],
\label{boxMink}
\end{align}
and
\begin{align}
\Box\langle|\varphi(x)|^2\rangle_b ^{(q,\alpha_0)}&=-\frac{8m^D}{(2\pi)^{D/2}}\left\{\sideset{}{'}\sum_{k=1}^{[q/2]}\cos(2k\pi\alpha_0)
\left[4m^2\left( r^2 s_k^4 +z^2\right)f_{\frac{D+2}{2}}\left(\gamma_{k,z}\right)-(2s_k^2+1)f_{\frac{D}{2}}(\gamma_{k,z})\right]\right.\nonumber\\
&-\left.
\frac{q}{\pi}\int_0^\infty dy \frac{h(q,\alpha_0,2y)}{\cosh(2qy)-\cos(q\pi)}\left[4m^2(r^2c_y^4 +z^2)\, f_{\frac{D+2}{2}}(\gamma_{y,z})-
(2c_y^2+1)f_{\frac{D}{2}}(\gamma_{y,z})\right]\right\}.
\end{align}
The above d'Alembertian operators are calculated from the Eqs. \eqref{fieldsquaredMink} and \eqref{fieldsquared}, respectively.

In the geometry under consideration, the differential operators $\nabla_r\nabla_r, \nabla_\phi\nabla_\phi$, $\nabla_z\nabla_z$
and $\nabla_r\nabla_z$ present contributions when acting on the VEV of the field squared.
In particular, for the azimuthal contribution, we shall use the expression \eqref{Wb}. After some intermediate steps,
we arrive at the summation below
\begin{equation}
S(q,\alpha,s)=\sum_{n=-\infty}^\infty q^2(n+\alpha)^2I_{q|n+\alpha|}(s),
\end{equation}
with $s=urr'$. We can use the following differential operator obeyed by the modified Bessel function
to evaluate the above summation:
\begin{equation}
S(q,\alpha,s)=\left(s^2\frac{d^2}{ds^2}+s\frac{d}{ds}-s^2\right)\sum_{n=-\infty}^\infty I_{q|n+\alpha|}(s),
\end{equation}
with \cite{Braganca2015}
\begin{eqnarray}
\sum_{n=-\infty}^\infty I_{q|n+\alpha|}(s)=\frac{e^{s}}{q}+\frac{2}{q}\sideset{}{'}\sum_{k=1}^{[q/2]}\cos(2k\pi\alpha_0)
e^{s\cos(2k\pi/q)}-\frac{2}{\pi}\int_0^\infty dy\frac{h(q,\alpha_0,2y)e^{-s\cosh(2y)}}{\cosh(2qy)-\cos(q\pi)}.
\end{eqnarray}

After long but straightforward calculations, we can decompose the boundary induced part of the energy-momentum tensor as
\begin{equation}
\langle T_\mu^\nu(x)\rangle_b=\langle T_\mu^\nu(x)\rangle_b^{(M)}+
\langle T_\mu^\nu(x)\rangle_b^{(q,\alpha_0)}.
\end{equation}
The first term in the r.h.s of the above equation is the VEV of the energy-momentum tensor induced by a boundary in Minkowski spacetime
in the absence of the magnetic flux. This contribution is obtained taking the $k=0$ term of \eqref{wb1} with the coefficient $1/2$
along with the Eqs. \eqref{fieldsquaredMink}
and \eqref{boxMink}. Then, the only nonzero contributions of the energy-momentum tensor in Minkowski spacetime in the
absence of the magnetic flux is written as
\begin{equation}
\langle T_\mu^\nu(x)\rangle_b^{(M)}=-\frac{2m^D}{(2\pi)^{D/2}}\left[4m^2z^2(4\xi-1)\, f_{\frac{D+2}{2}}(2m|z|)
+2(1-2\xi)f_{\frac{D}{2}}(2m|z|)\right],
\label{tensorMink}
\end{equation}
for the components $\mu=\nu=0,1,2,4, ..., D-1$.
Note that for a massless scalar field, the above equation reduces to
\begin{equation}
\langle T_\mu^\nu(x)\rangle_b^{(M)}=-\frac{4\Gamma(D/2)}{(4\pi)^{D/2}|z|^D}(D-1)(\xi-\xi_D),
\label{tensorMink_massless}
\end{equation}
where $\xi_D=\frac{(D-2)}{4(D-1)}$. For a conformally coupled massless scalar field, $\xi=\xi_D$, we
have that \eqref{tensorMink} vanishes.

The components of the
energy-momentum tensor induced by the boundary and the magnetic flux are given by the followings VEVs:
\begin{eqnarray}
\langle T_\mu^\nu(x)\rangle_b^{(q,\alpha_0)}&=&-\frac{4m^D}{(2\pi)^{D/2}}
\left[\sideset{}{'}\sum_{k=1}^{[q/2]}\cos(2k\pi\alpha_0)\, \mathcal{G}_\mu^\nu(2mr,2mz,s_k)\right.\nonumber\\
&&\left.-\frac{q}{\pi}\int_0^\infty dy \frac{h(q,\alpha_0,2y)}{\cosh(2qy)-\cos(q\pi)}\, \mathcal{G}_\mu^\nu(2mr,2mz,c_y)
\right],
\label{tensor}
\end{eqnarray}
 where we have defined the notation
 \begin{align}
 \mathcal{G}_0^0(u,v,\omega)&=(4\xi-1)(\omega^4 u^2+v^2)\,f_{\frac{D+2}{2}}(\gamma)+
 [1-(4\xi-1)(1+2\omega^2)]\,f_{\frac{D}{2}}(\gamma),\nonumber\\
 \mathcal{G}_1^1(u,v,\omega)&=(4\xi-1)v^2\,f_{\frac{D+2}{2}}(\gamma)+
 2[1-2\xi(1+\omega^2)]\,f_{\frac{D}{2}}(\gamma),\nonumber\\
  \mathcal{G}_2^2(u,v,\omega)&=[4\xi(\omega^4u^2+v^2)-\gamma^2]\,f_{\frac{D+2}{2}}(\gamma)+
 2[1-2\xi(1+\omega^2)]\,f_{\frac{D}{2}}(\gamma),\nonumber\\
   \mathcal{G}_3^3(u,v,\omega)&=(4\xi-1)\omega^2[\omega^2u^2\,f_{\frac{D+2}{2}}(\gamma)-
 2\,f_{\frac{D}{2}}(\gamma)],\nonumber\\
    \mathcal{G}_1^3(u,v,\omega)&=-(4\xi-1)\omega^2uv\,f_{\frac{D+2}{2}}(\gamma).
 \end{align}
We note that the energy-momentum tensor is an even function of $\alpha_0$. In the above notation, the indices $0,1,2,3$ correspond to the coordinates $t,r,\varphi,z$. As a consequence of boost invariance of our system
along the directions $x^j,j=4,...,D-1$, we have the relation
 $\langle T_j^j(x)\rangle_b=\langle T_0^0(x)\rangle_b$, for the components (no summation over $j$) with $j=4,...,D-1$. From the above expressions, we note that $\langle T_0^0(x)\rangle_b\neq \langle T_3^3(x)\rangle_b$. This is a consequence of the lost of invariance
along the string axis due the presence of the boundary.

Considering a massless scalar field, the VEVs of the energy-momentum tensor induced by the boundary and
the magnetic flux are given by
\begin{eqnarray}
\langle T_\mu^\nu(x)\rangle_b^{(q,\alpha_0)}&=&-\frac{2\Gamma(D/2)}{(4\pi)^{D/2}}\left[\sideset{}{'}\sum_{k=1}^{[q/2]}
\frac{\cos(2k\pi\alpha_0)\,\mathcal{F}_\mu^\nu(r,z,s_k)}{\left(r^2s_k^2+z^2\right)^{D/2+1}}\right.\nonumber\\
&&\left.-\frac{q}{\pi}\int_0^\infty dy \frac{h(q,\alpha_0,2y)}{\cosh(2qy)-\cos(q\pi)}\, \frac{\mathcal{F}_\mu^\nu(r,z,c_y)}
{\left(r^2c_y^2+z^2\right)^{D/2+1}}\right],
\label{tensor_massless}
\end{eqnarray}
with the notation
\begin{align}
\mathcal{F}_0^0(r,z,\omega)&=\left\{(4\xi-1)[(D-2)\omega^2-1]+1\right\}r^2\omega^2
[(4\xi-1)(D-1-2\omega^2)+1]z^2,\nonumber\\
\mathcal{F}_1^1(r,z,\omega)&=[2-4\xi(1+\omega^2)]r^2\omega^2+[4\xi(D-1-\omega^2)-D+2]z^2,\nonumber\\
\mathcal{F}_2^2(r,z,\omega)&=\left\{4\xi[(D-1)\omega^2-1]-D+2\right\}r^2\omega^2
+[4\xi(D-1-\omega^2)-D+2]z^2,\nonumber\\
\mathcal{F}_3^3(r,z,\omega)&=(4\xi-1)\omega^2[(D-2)r^2\omega^2-2z^2],\nonumber\\
\mathcal{F}_1^3(r,z,\omega)&=-D(4\xi-1)rz\omega^2.
\end{align}

\begin{figure}[h]
	\centering
	{\includegraphics[width=0.48\textwidth]{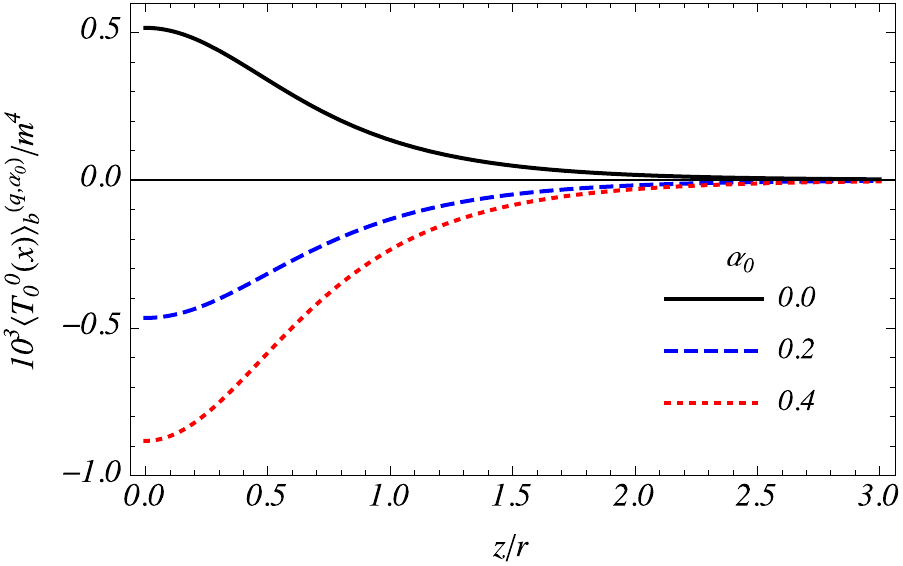}}
	\hfill
	{\includegraphics[width=0.49\textwidth]{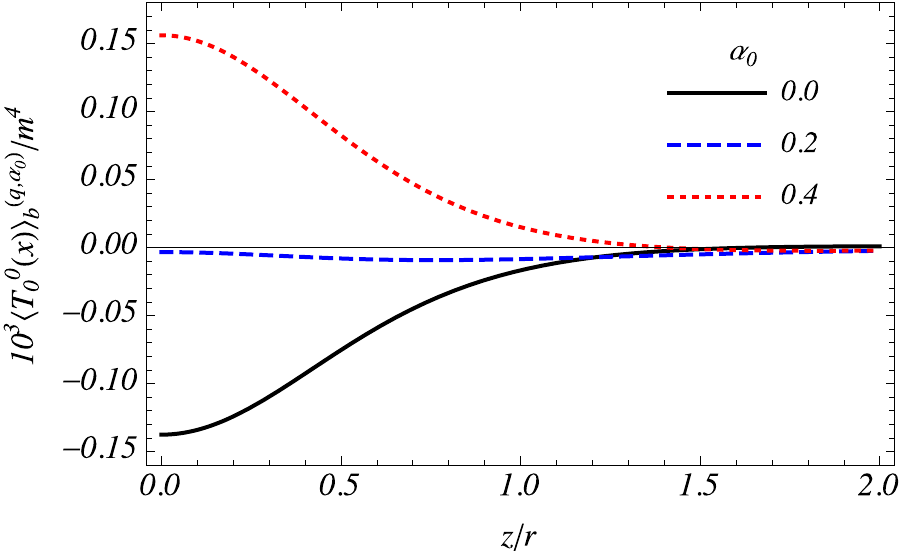}}
	\caption{VEV of the energy-density induced by the boundary and the magnetic flux as function of $mz$, considering
	different values of the parameter $\alpha_0$ for a minimally (left plot) and conformally (right plot)coulped scalar field.
	In addition, the plots are considered for $D=4$, $mr=0.75$ and $q=1.5$.}
	\label{fig03}
\end{figure}

The Fig. \ref{fig03} shows the behavior of the VEV of the energy density induced by both the boundary, conical geometry and
magnetic flux as function of $z/r$, for different values of $\alpha_0$. We note that the $\langle T_\mu^\nu(x)\rangle_b^{(q,\alpha_0)}$
depend crucially on the curvature coupling. In addition, similarly to the azimuthal current density, the energy density is finite at the origin for points
outside the boundary if $q|\alpha_0|>1$. In the case where $q|\alpha_0|<1$, the energy density diverge on the origin as
$\langle T_\mu^\nu(x)\rangle_b^{(q,\alpha_0)}\propto r^{-2(1-q|\alpha_0|)}$.

The energy-momentum tensor induced by the boundary presents a off-diagonal component. Consequently, from the covariant conservation
condition, $\nabla_\mu \langle T_\nu^\mu (x)\rangle_b=0$, we found the following non-trivial differential equations
\begin{equation}
\partial_r\left(r\langle T_1^1(x)\rangle_b\right)+r\partial_z\langle T_1^3(x)\rangle_b=\langle T_2^2(x)\rangle_b
\end{equation}
and
\begin{equation}
\partial_z \langle T_3^3(x)\rangle_b=-\frac{1}{r}\partial_r\left(r\langle T_1^3(x)\rangle_b\right).
\end{equation}
It is possible to check that the previous expressions found for the energy-momentum tensor obey the above relations.
In addition, the VEV of the energy-momentum tensor obey the trace relation
\begin{equation}
\langle T_\mu^\mu(x)\rangle_b=2(D-1)(\xi-\xi_D)\nabla_\mu\nabla^\mu \langle |\varphi(x)|^2\rangle_b
+2m^2\langle |\varphi(x)|^2\rangle_b.
\end{equation}
Note that the energy-momentum tensor is traceless for a massless conformally coupled field ($\xi=\xi_D$).

The component $\langle T_3^3(x)\rangle_b$ at $z=0$ determines the normal vacuum force on the boundary. This contribution is finite outside
the string axis and is written as
\begin{align}
\langle T_3^3(x) \rangle_b^{(z=0)}&=\frac{4m^D(1-4\xi)}{(2\pi)^{D/2}}\left[\sideset{}{'}\sum_{k=1}^{[q/2]}s^2_k\cos(2k\pi\alpha_0)\, 
\mathcal{G}_3(2mrs_k)-\frac{q}{\pi}\int_0^\infty dy \frac{h(q,\alpha_0,2y)\mathcal{G}_3(2mrc_y)}{\cosh(2qy)-\cos(q\pi)}\,\right],
\label{T3}
\end{align}
with the notation
\begin{equation}
\mathcal{G}_3(v)=v^2\, f_{\frac{D+2}{2}}(v)-2f_{\frac{D}{2}}(v).
\end{equation}
We have that \eqref{T3} is equivalent to the effective pressure on the boundary, i.e., $\mathcal{P}=\langle T_3^3(x) \rangle_b^{(z=0)}$, and presents
a dependence on the curvature coupling parameter in the form of the factor $(1-4\xi)$.
\begin{figure}[h]
	\centering
	{\includegraphics[width=0.49\textwidth]{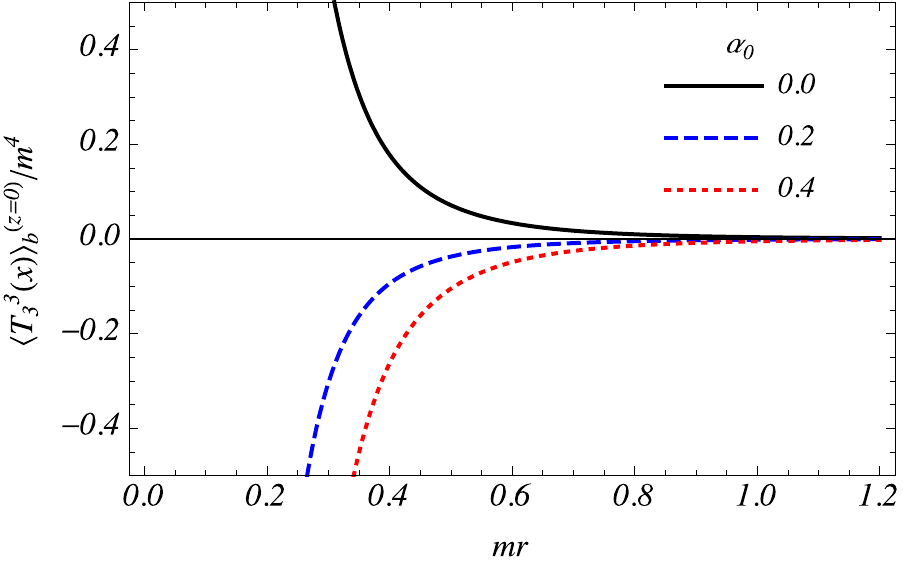}}
	\hfill
	{\includegraphics[width=0.49\textwidth]{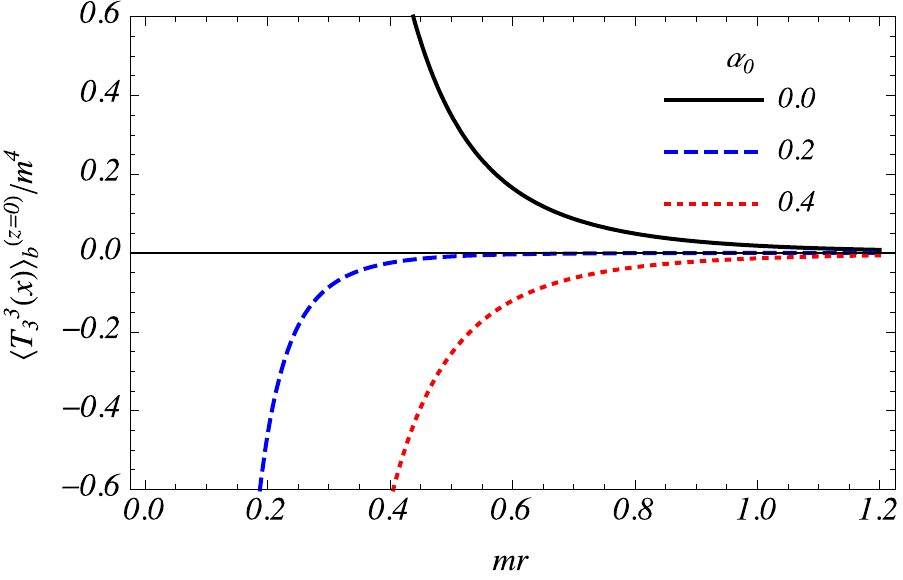}}
	\caption{VEV of the vacuum force on the boundary as function of $mr$ for different values of $\alpha_0$ and a minimally coupled scalar field considering $q=1.5$ (left plot) and $q=2.5$ (right plot). In both plots we also have $D=4$.}
	\label{fig04}
\end{figure}

In Fig. \ref{fig04} we show the behavior of the normal vacuum force \eqref{T3} as function of $mr$ considering different values of the parameter $\alpha_0$ for $q=1.5$ (left plot) and $q=2.5$ (right plot).
In the absence of the magnetic flux, $\alpha_0=0$, we have that the effective pressure is always positive and is consequence only of the conical topology of the spacetime. However, when we take into account the presence of a magnetic flux along the string axis, the effective pressure can assume positive or negative values. Although not exhibited in the graph, the pressure in the case of absence of magnetic flux is also positive for conformally coupled field. In addition, we note that intensity of the vacuum force
increases with the angle deficit $q$.\footnote{Although there is no physical reason to adopted specific values for $q$, the values assumed in the Fig. \ref{fig04}, $q=1.5$ and $2.5$, are conveniently chosen to exhibit the behavior of the normal force with $q$.} 

\section{Conclusions}
\label{conc}
In the present paper, we have investigate the vacuum bosonic current and polarization associated with a quantum charged scalar field
in a higher-dimensional cosmic string spacetime considering the presence of a flat boundary orthogonal to the string.
We also have considered the presence of a magnetic flux running along the string axis and that the quantum field obeys the
Dirichlet or Neumann boundary condition on the boundary. As the first step of our analysis, we evaluate the Wightman function
and we found a closed form of it for general values of the parameter that codifies the presence of the conical defect, $q$.
The Wightman function could be decomposed into two contributions, one in the spacetime time of a cosmic string in the absence
of the planar boundary and another one induced by the boundary, Eqs. \eqref{wcs1} and \eqref{wb1}, respectively. As the vacuum induced observable in the absence of the boundary have been investigated in the literature by several authors, here we were concerned only
in the analysis of the planar boundary effects.

The induced bosonic current was the first physical quantity that we have developed. The only nonzero component of the bosonic current is 
the azimuthal one and the contribution induced by the boundary is given by the Eq. \eqref{jphi_b} which can be positive or negative
depending on the boundary condition obeyed by the scalar quantum field. In addition, the azimuthal current induced by the boundary
is an odd function of $\alpha_0$ and is finite on the string for points outside the boundary if $q|\alpha_0|>1$. For the case where $q|\alpha_0|<1$
the azimuthal current is divergent on the string and the leading divergence term is given by \eqref{div.J}. We also have calculated the
boundary induced part of the azimuthal current considering large distances from the boundary and in the limit of a massless scalar field, expressed by Eqs. \eqref{jphi_b1} and \eqref{jphi_b2}, respectively. In the Fig. \ref{fig01} we have exhibited the behavior of the azimuthal current induced by the boundary as function of $mr$ and $z/r$ considering $D=4$, where we note that the azimuthal current density is intensified as the parameter $q$ increases. In the addition, we have shown that depending on the boundary condition adopted, the total azimuthal current density
can be canceled or intensified on the boundary.

Our next step was to develop the calculation of the vacuum polarization considering the VEV of the field squared and
the energy momentum tensor. The field squared induced by the boundary could be decomposed into two contributions:
one contribution induced by a planar boundary in Minkowiski spacetime, Eq. \eqref{fieldsquaredMink}, and another
one induced by the conical defect and the magnetic flux, Eq. \eqref{fieldsquared}. The first is independent of the
magnetic flux and the radial coordinate and the latter is an even function of $\alpha_0$. For the boundary induced part of
the field squared we have considered some special cases. For the regions near the boundary the leading contribution
comes from the Eq. \eqref{fieldsquaredMink} and is given by \eqref{FieldSmallZ}. In the Eq. \eqref{fieldsquared_massless}
we have the boundary part of the azimuthal current induced by the conical defect and the magnetic flux considering
a massless scalar field. Also, this part of the azimuthal current were considered taking into account large distances from the boundary,
Eq. \eqref{fieldLargeZ}, which presents an exponential decay with $z/r$. The behavior of the field squared induced by the magnetic
flux and the conical defect is shown in the Fig. \ref{fig02} as function of $z/r$ and $mr$ considering $D=4$.

We also developed the analysis of the another quantity that characterizes the quantum vacuum state: the VEV of the energy-momentum
tensor. As the previous physical quantities, we developed only the analysis of the boundary effects of the energy-momentum tensor,
which could be decomposed into a contribution due a flat boundary in Minkowisk spacetime, Eq. \eqref{tensorMink},
and another one induced by the magnetic flux and the conical defect, Eq. \eqref{tensor}. For the first, the only nonzero contributions
are the components $\mu=\nu=0,1,2,4, ..., D-1$. This part of the energy-momentum tensor also was calculated considering a massless
scalar field, Eq. \eqref{tensorMink_massless}. The contribution of the energy-momentum tensor induced by the magnetic flux and the
conical defect is an even function of $\alpha_0$. We have found the relation $\langle T_j^j(x)\rangle_b=\langle T_0^0(x)\rangle_b$,
for the components along the extra dimensions. This is a directly consequence of the boost invariance along these directions. However,
this invariance is lost along the $z$-direction due the presence of the boundary, consequently, $\langle T_0^0(x)\rangle_b\neq \langle T_3^3(x)\rangle_b$.
We also have considered the VEV of the energy-momentum tensor induced by the boundary for the case of a massless scalar field, which is
given by the Eq. \eqref{tensor_massless}. In the Fig. \ref{fig03} we have plotted the behavior of boundary part of the energy density
induced by the magnetic flux and conical defect as function of $z/r$ considering $D=4$, where we note that the energy density depends crucially of the curvature coupling. In addition,
the energy density in finite on the string for points outside the boundary only if $q|\alpha_0|>1$ and diverges as
$\langle T_\mu^\nu(x)\rangle_b^{(q,\alpha_0)}\propto r^{-2(1-q|\alpha_0|)}$ at the origin if $q|\alpha_0|<1$.

To finish our analysis, we have evaluate the normal vacuum force on the boundary, which is determined by the component $\langle T_3^3\rangle_b$
at $z=0$. This contribution is given by the Eq. \eqref{T3} that presents a dependence on the curvature coupling through the factor $(1-4\xi)$.
And interesting characteristic of the normal vacuum force is that in the absence of the magnetic flux, this force has always positive
values for both minimally and conformally scalar fields, being consequence only of the conical topology of the two surface orthogonal to the string.
However, when the magnetic flux is present, the normal vacuum force can assume positive or negative values. The profile of the
normal vacuum force on the boundary is shown in the Fig. \ref{fig04} as function of $mr$ for $D=4$.

\section*{Acknowledgments}
E.R.B.M is partially supported by Conselho Nacional de Desenvolvimento Cient\'{\i}fico e Tecnol\'{o}gico - Brasil (CNPq) under grant No 301.783/2019-3.
E.A.F.B is grateful by the hospitality of the State University of the Tocantina Region of Maranh\~{a}o - Brazil.

\end{document}